\documentclass[lettersize,journal]{IEEEtran}
\usepackage{amsmath}
\usepackage{amsfonts}
\usepackage{amsthm}
\usepackage{amssymb}
\usepackage{bbm}

\usepackage{algorithmic}
\usepackage{algorithm}

\usepackage{array}
\usepackage[caption=false,font=normalsize,labelfont=sf,textfont=sf]{subfig}
\usepackage{textcomp}
\usepackage{stfloats}
\usepackage{url}
\usepackage{graphicx}
\usepackage{cite}

\newtheorem{remark}{Remark}

\begin{document}
\title{\huge Latency Fairness Optimization on Wireless Networks through Deep Reinforcement Learning}

\author{
\IEEEauthorblockN{\normalsize M. López-Sánchez, A. Villena-Rodríguez, G. Gómez, F. J. Martín-Vega, and M. C. Aguayo-Torres}
%
%
\thanks{Manuscript received November XX, 2021; revised XX. This work was supported by the European Fund for Regional Development (FEDER), Junta de Andaluc\'ia and the University of M\'alaga under the projects P18-RT-3175, P18-TP-3587, UMA-CEIATECH-06 and the postdoctoral grant Ref., DOC\_00265 (``Selecci\'on de personal investigador doctor convocado mediante resoluci\'on de 21 de mayo de 2020'', PAIDI 2020).}
\thanks{The authors are with the Communications and Signal Processing Lab, Instituto Universitario de Investigaci\'on en Telecomunicaci\'on (TELMA), Universidad de M\'alaga, CEI Andaluc\'ia TECH, ETSI Telecomunicaci\'on, Bulevar Louis Pasteur 35, 29010 M\'alaga (Spain) (e-mail: \{mls, avr, ggomez, fjmvega, aguayo\}@ic.uma.es)}
}

\maketitle



\maketitle

\begin{abstract} 
In this paper we propose a novel deep reinforcement learning framework to maximize user fairness in terms of delay. To this end, we devise a new version of the modified largest weighted delay first (M-LWDF) algorithm, which is called \mbox{$\beta$-M-LWDF}, aiming to fulfill an appropriate balance between user fairness and average delay. This balance is defined as a feasible region on the cumulative distribution function (CDF) of the user delay that allows to identify \textit{unfair} states, \textit{feasible-fair} states and \textit{over-fair} states. Simulation results reveal that our proposed framework outperforms traditional resource allocation techniques in terms of latency fairness and average delay.
\end{abstract}

\begin{IEEEkeywords}
Scheduling, latency, 5G, Reinforcement Learning, Deep Learning
\end{IEEEkeywords}

\section{Introduction}
%
\IEEEPARstart{L}{ow} latency has been one of the key features that next wireless systems, such as 5G and beyond networks, are designed to fulfill in a wide range of scenarios. Nevertheless, this requisite is probably one of the most challenging from the quality of service (QoS) provision point of view. 

On the one hand, achieving small block error rates with latencies of the order of few milliseconds requires the specific design of the radio interface functions, specially the radio scheduler. On the other hand, when the system is handling a traffic load close to its capacity, the packet delay can greatly increase leading to packet loss and an unacceptable latency. These two aspects pose an interesting challenge to the design of the scheduling algorithms that must minimize the average delay while providing user fairness. 

%
In this sense, the use of hybrid control systems based on machine learning (ML) is considered as a promising approach that enables intelligent and adaptive decision making to the variant conditions of the network. 
%
In \cite{Pedersen18}, the essential role played by the radio scheduler in 5G is tackled. From a more theoretical point of view, reinforcement learning (RL) is presented in \cite{Sutton14} as a promising solution that allows dealing with the dynamic radio environment and learning mapping functions over time. For instance, research advances described in \cite{Comsa19} and \cite{Comsa19bis} analyze the user fairness in terms of throughput for a modified proportional fair (PF) strategy, finding a balance between the system spectral efficiency and the fairness among users. However, to the best of the author's knowledge, user fairness in terms of delay has not been addressed yet. This is a critical issue in the emerging low latency services expected for 5G and beyond. Therefore, it is meaningful to investigate this issue and propose smart resource allocation algorithms that keep a balance between user fairness and system performance in terms of delay. 

In this paper, an intelligent controller is proposed to customize a new version of the modified largest weighted delay first (M-LWDF) scheduling strategy \cite{LDF} at each Transmission Time Interval (TTI). In particular, we propose a novel utility function called $\beta$-M-LWDF that is able to adjust dynamically the experienced delay by means of a parameter $\beta$ that is managed by the controller. It is important to remark that other scheduling algorithms such as largest delay first (LDF) \cite{LDF}, PF and M-LWDF can be viewed as special cases of our proposed utility function for $\beta \to \infty$, $\beta = 0$ and $\beta = 1$, respectively. Besides, with our proposed ML scheme, the $\beta$ parameter is selected at each TTI to achieve an appropriate balance between user fairness and system performance in terms of delay. Such a balance is defined as a feasible-fair region on the cumulative distribution function (CDF) of the normalized delay.  This allows us to identify \textit{unfair} states, where the difference of the delay between different users is excessive; \textit{over-fair} states, where the difference is small, but the average delay increases; and \textit{feasible-fair} states which reach an appropriate balance between latency fairness and average delay.  
To cope with the high  complexity of this approach, the controller implements a RL algorithm known as deep Q-learning (DQL) as introduced in \cite{MNIH15}, which makes use of a neural network to estimate the Q-Function that approximates the best decisions at each TTI. 


\section{System Model}
\label{sec:system_model}
We consider the downlink (DL) of the 5G NR (New Radio) system, which is based on Orthogonal Frequency Division Multiple Access (OFDMA). The available bandwidth is divided into equal resource blocks (RBs), where one RB consists on 12 sub-carriers and it represents the minimum resource unit that can be assigned in the frequency domain. The sub-carrier spacing (SCS) can be expressed as $\Delta f = 2^\mu 15$ kHz, where $\mu$ is the numerology and it ranges from $0$ to $3$ for data channels.  The time domain resources are divided in slots, also named TTIs. 
It is considered adaptive modulation and coding (AMC), and thus the transmission rate, i.e., the modulation and coding scheme (MCS), is chosen to maximize the throughput while guaranteeing a block error rate (BLER) below a target value, $\mathrm{BLER}_T = 10 \%$ \cite{Martin21}. The set of MCSs are taken from 5G specs (Table 5.1.3.1-1 of 38.214 v16.7.0). 
Let $\mathcal{B} = \{b_1, b_2,…, b_N\}$ be the set of RBs for a given bandwidth, where $N$ represents the total number of RBs. Additionally, let $\mathcal{U} = \{u_1, u_2,…, u_M\}$ be the set of users to be scheduled. 
%
%
A constant bit rate (CBR) traffic model is assumed for each user where a packet sizes of $S_\mathrm{CBR}$ bytes reach the buffer every $T_\mathrm{CBR}$ seconds.  
It is assumed that the packets are stored in a unbounded buffer and the packets are segmented according to the transport block (TB) size that is determined by the scheduler and the selected MCS. 
%
%
The average signal-to-noise ratio (SNR) of each user, $\bar{\gamma_u}$, is drawn randomly according to a Log-Normal distribution with expected value $\mu_\gamma$ and standard deviation $\sigma_\gamma$. In addition, the multi-path fading follows a realistic Tapped Delay Line Channel (TDL-A) model where the channel complex samples,  
$H_{u,k}[n]$, are correlated in time and frequency domains. Thus, the instantaneous SNR of user $u$ at RB $k$ and TTI $n$ is computed as $\gamma_{u,k}[n] = \bar{\gamma_u} |H_{u,k}[n]|^2$. 


\subsection{Scheduling algorithms}
\label{sec:alg}
The aim of the scheduler is to allocate each RB $b_i \in B$ to a particular user $u_i \in U$ at each TTI in order to meet a predefined QoS requirement. We have evaluated four different scheduling policies:

\begin{enumerate}
    \item 
    The well known Proportional Fair (PF) uses the following utility function at each TTI $n$, $U[n]$, as decision criterion to assign the user $\widehat u$ to the RB $\widehat k$ on the time instant $n$:
\begin{equation}
\label{eq:PF}
\small
\left\{ {\widehat u[n],\widehat k[n]} \right\} = \arg \mathop {\max }\limits_{u,k} \left\{ {\frac{{r_{u,k} [n]}}{{\overline {r_u }[n] }}} \right\},
\end{equation}
\noindent 
where $r_{u,k}[n]$ represents the potential rate for user $u$ on the RB $k$ and TTI $n$, and $\overline {r_u }[n]$ is a weighted moving average of the data rate values reached in previous TTIs, which is computed as \cite{Musleh15}
\begin{equation}
\label{eq:r_PF}
\small
\overline {r_u } \left[n \right] = \left( {1 - \frac{1}{{T_{PF} }}} \right)\overline {r_u } \left[ {n - 1} \right] + \frac{1}{{T_{PF} }}r_u \left[ {n - 1} \right],
\end{equation}
\noindent 
where $\overline {r_u }[n-1]$ represents the weighted average data rate reached up to the TTI $n-1$, $r_u [n-1]$ is the data rate reached in TTI $n-1$, and $T_{PF}> 0$ is the average window size. 
%

    \item 
LDF algorithm \cite{LDF} is adaptive to the delay, providing the turn to the user that has been suffering the largest delay in its queue, as follows:
\begin{equation}
\small
\label{eq:LDF}
\widehat u[n] = \arg \mathop {\max}\limits_{u} \left\{ { W_u [n] } \right\}.
\end{equation}
\noindent 
where $W_u [n]$ is the delay experienced in the queues, which is computed as the sum of the delays of every packet that is stored in the queue. 
%
    \item 
The algorithm M-LWDF considers the waiting time in the queues, the instantaneous capacity of the channels and a parameter related to the delay tolerance \cite{Andrews2002}:
\begin{equation}
\small
\label{eq:M-LWDF}
\left\{ {\widehat u[n],\widehat k[n]} \right\} = \arg \mathop {\max} \limits_{u,k} \left\{ {g_u [n] \cdot W_u [n] \cdot r_{u,k} [n]} \right\},
\end{equation}
\noindent
where $g_u [n] \in [0, \infty)$ is a QoS factor with a value of $g_u [n] = a_u / \overline {r_u } [n]$, with $\overline {r_u } [n]$ as defined in \eqref{eq:r_PF} and $a_u  =  - \log \left( {\delta _u } \right) \cdot T_u$, being $\delta_u \in [0, 1]$ the desired probability of fulfilling the delay requirement $T_u$ \cite{Andrews2002}.


    \item 
We propose a new version of the M-LWDF, called $\beta$-M-LWDF, in which the impact of the delay term can be adjusted by a parameter $\beta$ that is selected by the intelligent controller at each TTI:
\begin{equation}
\small
\label{eq:bmlwdf}
\left\{ {\widehat u[n],\widehat k[n]} \right\} = \arg \mathop {\max}\limits_{u,k} \left\{ {g_u  \cdot W_u^{\beta} [n] \cdot r_{u,k} [n]} \right\}.
\end{equation}
\end{enumerate}
\begin{remark}
\label{rem:special cases}
The utility functions of PF, LDF and M-LWDF that are given with \eqref{eq:PF}, \eqref{eq:LDF} and \eqref{eq:M-LWDF} are special cases of the proposed $\beta$-M-LWDF with $\beta = 0$, $\beta \to \infty$ and $\beta = 1$, respectively. 
\end{remark}
\begin{proof}
It is clear that $\beta$-M-LWDF reduces to M-LWDF algorithm when $\beta=1$. For $\beta=0$, $\beta$-M-LWDF reduces to PF as long as the parameter $a_u$ is equal for all users (as considered in this work). The proof for the case of LDF when $\beta \to \infty$ is given in Appendix \ref{proof:remark}. 
\end{proof}
\vspace{-3mm}
\begin{remark}
\label{rem:beta behaviour}
As it is shown in \cite{LDF}, LDF provides the highest fairness in terms of delay while sacrificing system performance (i.e., average delay), whereas M-LWDF achieves a higher system performance at the expense of fairness. This is due to the fact that LDF only accounts for the user delay whereas M-LWDF also accounts for the instantaneous rate. In view of remark \ref{rem:special cases},  this involves that increasing $\beta$ increases the fairness while decreasing $\beta$ reduces the fairness in terms of delay. 
\end{remark}
%
%
\section{Proposed RL Framework}
\label{sec:framework}
\subsection{RL Framework Description}
\begin{figure}[t]
\centering
\includegraphics[width=0.85\columnwidth]{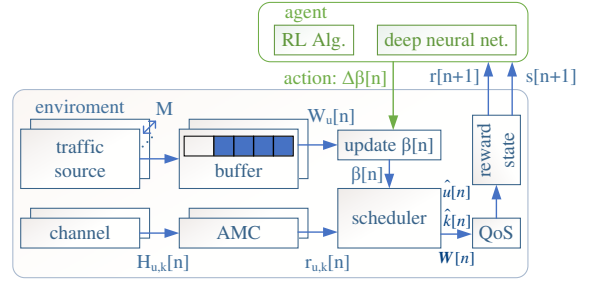}
\caption{Proposed RL framework}
\label{fig:framework}
\end{figure}
Our proposed RL framework is shown in Fig. \ref{fig:framework}. At TTI $n$, the \textit{channel} block generates the multi-path fading samples. 
Then the instantaneous SNR, $\gamma_{u,k}[n]$, is computed. This SNR is used by the \textit{AMC} block to select the MCS that maximizes the potential rate, $r_{u,k}[n]$, while keeping a BLER below $\mathrm{BLER}_T$. On the other hand, packets are generated by the \textit{traffic source} and stored in the \textit{buffer} queue. This block performs segmentation of the stored packets according to the TB size determined by the \textit{scheduler} and it computes the queue delay of each user $W_u[n]$. 
Afterwards, the scheduler updates its $\beta[n]$ parameter according to the actions made by the agent and it allocates resources based on the proposed $\beta$-M-LWDF utility function. Once the scheduler has performed the resource allocation, the \textit{QoS} block estimates the empirical CDF of the normalized user delay, which is used to compute the reward and state variables. Those variables are used by the agent to make the next action in the following TTI. This action involves determining the step $\Delta \beta [n] \in [-1,+1]$ to increase or decrease the beta parameter as $\beta[n] = \beta[n-1] + \Delta \beta [n]$.
%
%
\vspace{-3mm}
\subsection{Latency Fairness Criteria}
The Next Generation Mobile Networks (NGMN) Alliance defines an impartiality requirement in terms of throughput. This criteria is defined so that a system is seen as fair if at least the $(100-x)\%$ of active users reach at least $x\%$ of the normalized user data rate \cite{NGNM}. 
%
%

In this work we propose a RL framework intended to fulfill a predefined user fairness criteria in terms of latency. However, up to date there is no standardized fairness requirement in terms of delay. For that reason, in this paper we propose an impartiality requirement with the following $3$ desired features: i) half of the users have a delay smaller than the average; ii) there are no users with a delay $50\%$ higher than the average; and iii) there are no users with a delay $50\%$ smaller than the average. 
These $3$ desired features lead to a CDF requirement as an straight line that passes through the points $(0.5, 0)$ and $(1.5, 1)$ as follows
\vspace{-3mm}
\begin{equation}
\small
\label{eq:CDF requirement}
y = f_R(w) = 
    \begin{cases}
        1 & w > 1.5 \\
        w - 0.5 & 0.5 \leq w \leq 1.5 \\
        0 & w < 0.5
    \end{cases}
\end{equation}
\noindent 
where $w = f^{-1}_R (y) = y + 0.5$ stands for the inverse CDF requirement. 
It can be noticed that such a CDF requirement  can be also read as $c\%$ of the users to be below the $(50-c)\%$ of the normalized user delay, $\tilde{W}_u[n]$, which is defined as
%
\begin{equation}
\small
\label{eq:normalized delay}
\tilde{W}_u[n] = \frac{W_u[n]}{\frac{1}{M} \sum_{u^\prime = 1}^M W_{u^\prime}[n]} 
\end{equation}

As it can be noticed, $\tilde{W}_u[n]=1$ involves that the delay of the user at TTI $n$ is equal to the average user delay. In a perfectly fair system all the users would have the same delay, and thus, all of them would have a unitary normalized delay.

Nevertheless, as it is discussed in \cite{Sadr09, LDF} there is a trade-off between fairness and system performance. This involves that increasing the delay fairness is at the expense of increasing the average delay. For that reason, our proposed CDF requirement aims at identifying those cases where the fairness severely degrades the average delay, and those cases where the difference of delay between users is excessive. The former case is labeled as \textit{over-fair (OF)} and the latter is labeled as \textit{unfair (UF)}. Those cases that do not fall on the two aforementioned cases are labeled \textit{feasible-fair (FF)}.  
The goal of the agent is to maximize the number of TTIs where the system is in FF state since this case leads to an appropriate balance between fairness and average delay. 

The empirical CDF of the normalized delay at TTI $n$ is expressed below
\vspace{-3mm}
\begin{equation}
\small
    \label{eq:empirical CDF}
    \hat{F}_{\tilde{W}[n]}(w) = \frac{1}{M} \sum_{u=1}^M \mathbbm{1} \left( \tilde{W}_u[n] \leq w  \right) 
\end{equation}
%
\noindent
where $\tilde{W}[n]$ represents a randomly chosen normalized user delay at TTI $n$. $ \mathbbm{1}\left( \mathcal{E}  \right)$ stands for the indicator function, which is $1$ if the event $\mathcal{E}$ is true and $0$ otherwise. Let us represent the sorted vector of normalized delays as $\mathbf{W}^\uparrow[n]=\left(W^\uparrow_1[n], .., W^\uparrow_M[n]\right)$, where $W^\uparrow_j[n]$ is the $j$-th delay in ascending order. Then, if we evaluate the empirical CDF from \eqref{eq:empirical CDF} on $\mathbf{W}^\uparrow[n]$ we get $M$ equally spaced samples between $0$ and $1$ as follows $\mathbf{Y}[n]=(Y_1[n],..,Y_M[n]) = \hat{F}_{\tilde{W}[n]}\left(\mathbf{W}^\uparrow\right[n])$ where $Y_j[n] = \frac{j}{M} = \hat{F}_{\tilde{W}[n]}\left(W^\uparrow_j[n]\right)$. Therefore, we can obtain a required normalized delay, $w^{(R)}_j$, for each of the sorted users as $w^{(R)}_j=f^{-1}_R \left(\frac{j}{M} \right) = \frac{j}{M} + 0.5$ with $j\in [1,M] \subset \mathbb{N}$. A given sorted user $j$ is said to fulfill the delay requirement if $\Delta W_j[n] = W^\uparrow_j[n] - w^{(R)}_j \leq \xi$, where $\xi$ is a confidence factor.

\begin{algorithm}
\small
   \hspace*{\algorithmicindent} \textbf{Input:} $\mathbf{\tilde{W}}[n]=\left(\tilde{W}_1[n], .., \tilde{W}_M[n] \right)$ \\
   \hspace*{\algorithmicindent} \textbf{Output:} $\mathcal{C}[n]$, fairness case  
\begin{algorithmic}[1]   
\STATE   Sort the vector of normalized delays, $\mathbf{\tilde{W}}[n]$, to get $\mathbf{W}^\uparrow[n]$ 
   \IF {$ \left[ \sum_{j=1}^{\lceil \lambda M \rceil} \mathbbm{1} \left( W^\uparrow_j > w^{(R)}_j + \xi \right) \right] > 1$ }
   {
      \STATE $\mathcal{C}[n] = \mathrm{OF}$ 
   }
   \ELSIF {$ \left[ \sum_{j= \lceil \lambda M \rceil +1}^{M - \lceil \psi M \rceil} \mathbbm{1} \left( W^\uparrow_j > w^{(R)}_j + \xi \right) \right] > 1$ }
   {
      \STATE $\mathcal{C}[n] = \mathrm{UF}$  
   }
   \ELSE
   {
       \STATE $\mathcal{C}[n] = \mathrm{FF}$  
   }
   \ENDIF
\end{algorithmic}
\caption{{Selection of fairness cases}}
\label{alg:case selection}
\end{algorithm}

Algorithm \ref{alg:case selection} is used to identify in which of the three cases $\mathcal{C}[n] \in \{\mathrm{OF}, \mathrm{UF}, \mathrm{FF}\}$ is the system working at a given TTI, $n$. It can be observed from line $2$ that the system is labeled as OF if any of the $\lceil \lambda M \rceil$ best users (i.e., users with smallest delay) have a normalized delay greater than the required delay plus the confidence factor. The parameter $\lambda \in [0, 1]$ represents the percentage of best users. An OF situation means that the best users have a delay closer to the average delay, which increases the fairness. 
However, this happens at the expense of a significant reduction on the average delay. 

From line $4$ of algorithm \ref{alg:case selection} it is observed that the system is labeled as UF if any of the worst users do not fulfill the delay requirement. The agent aims at avoiding this case in order to reduce the normalized delay of the worst users. 
As it can be seen, the $\lceil \psi M \rceil$ users with the greatest delays are not considered because they are treated as outliers, where $\psi \in [0, 1]$. This means that those users have such a poor channel condition that do not represent the conditions of the vast majority of users and thus, they are not considered. 
If all the users (excluding the outliers) fulfill the delay requirement (including the factor $\xi$), then the system is in FF case. 

\begin{figure}[t]
\centering
\includegraphics[width=0.71\columnwidth]{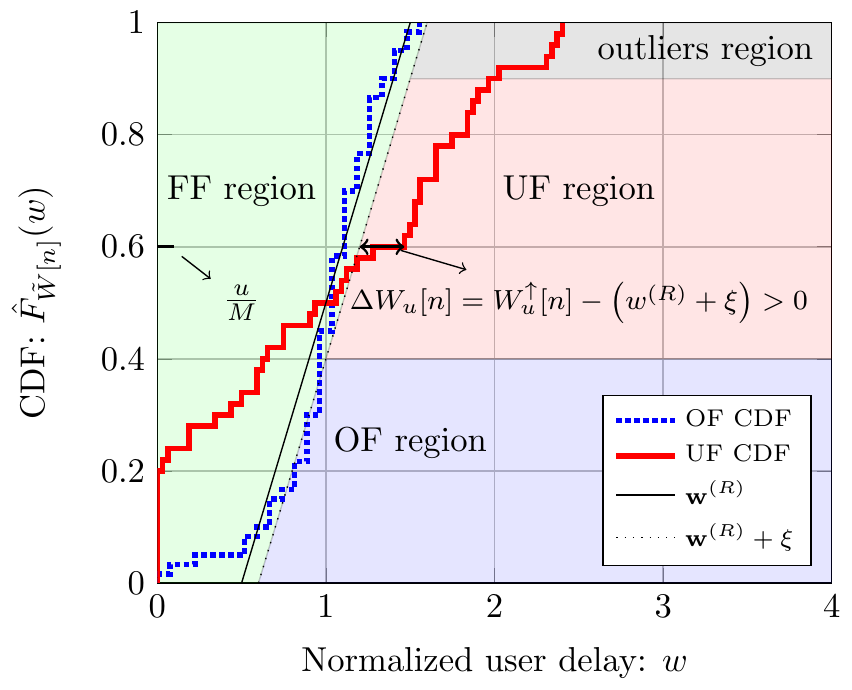}
\caption{Empirical CDF of the normalized delay for UF (red) and OF (blue) cases after defining $40\%$ of the users as best users (i.e., $\lambda=0.4$) and $10\%$ of the users as outliers ($\xi=0.1$) for a cell with $M=60$ users.}
\label{fig:cdf_req}
\end{figure}

Fig. \ref{fig:cdf_req} illustrates an example of the OF and UF cases by means of their estimated CDF of the normalized delay. As it is seen with algorithm \ref{alg:case selection}, the criteria to select the fairness case involves checking if any of the best or worst users do not fulfill the delay requirement. Nevertheless, this criteria is equivalent to check if the empirical CDF at some TTI $n$ falls within the OF, UF or FF regions respectively. Those regions are shown in Fig. \ref{fig:cdf_req} with blue, red and green colors. For instance, as it can be observed, the sorted user $u=36$, which makes the empirical CDF to reach $0.6$, has a normalized delay of $W_u^\uparrow=1.466$, which is greater than the related CDF requirement plus the confidence factor, i.e., $w^{(R)}+\xi = 1.1+0.1$. Since this user, and others that belong to the worst users set, do not fulfill the delay requirement, the system is labeled as UF in that case.

%
%
\vspace{-3mm}
\subsection{States and Actions}
Let $\mathcal{S}$ be the state space and let $s[n] \in \mathcal{S}$ be the instantaneous state at TTI $n$. The state $s[n]$ can be seen as a union of two sub-states $s[n] = \left(s_c[n], s_u[n] \right)$, the controllable ($s_c[n]$) and uncontrollable ($s_u[n]$) sub-states. 
The controllable sub-state is comprised of $s_c[n] = \left(\beta[n-1], {d}^{(\mathrm{inf})}[n], {d}^{(\mathrm{sup})}[n]\right)$ where ${d}^{(\mathrm{inf})}[n]$ and ${d}^{(\mathrm{sup})}[n]$ represent the distance of the normalized delay to the requirement for the best and worst users sets as described in algorithm \ref{alg:distance}.

\begin{algorithm}
\small
   \hspace*{\algorithmicindent} \textbf{Input:} $\mathbf{\Delta W}[n] = \left(\Delta W_1[n],.., \Delta W_M[n]\right)$  \\
   \hspace*{\algorithmicindent} \textbf{Output:} ${d}^{(\mathrm{inf})}[n], {d}^{(\mathrm{sup})}[n]$
\begin{algorithmic}[1]   

   \IF {$ \mathcal{C}[n] = \mathrm{OF}$ }
   {
      \STATE ${d}^{(\mathrm{inf})}[n] = \max \left(\mathbf{\Delta W}_{\left\{1:\lceil \lambda M \rceil \right\}}[n] \right)$  
      \STATE ${d}^{(\mathrm{sup})}[n] = \min \left(\mathbf{\Delta W}_{\{\lceil \lambda M \rceil+1:M\}}[n] \right)$ 
   }
   \ELSIF {$ \mathcal{C}[n] = \mathrm{UF}$ }
   {
      \STATE ${d}^{(\mathrm{inf})}[n] = \min \left(\mathbf{\Delta W}_{\{1:\lceil \lambda M \rceil\}}[n] \right)$  
      \STATE ${d}^{(\mathrm{sup})}[n] = \max \left(\mathbf{\Delta W}_{\{\lceil \lambda M \rceil+1:M\}}[n] \right)$  
   }
   \ELSE
   {
      \STATE ${d}^{(\mathrm{inf})}[n] = \min \left(\mathbf{\Delta W}_{\{1:\lceil \lambda M \rceil\}}[n] \right)$
      \STATE ${d}^{(\mathrm{sup})}[n] = \max \left(\mathbf{\Delta W}_{\{\lceil \lambda M \rceil+1:M\}}[n] \right)$  
   }
   \ENDIF
\end{algorithmic}
\caption{{Selection of state values}}
\label{alg:distance}
\end{algorithm}


For the OF case $d^{\rm (inf)}$ is the maximum of the distances of the best users while $d^{\rm (sup)}$ is the minimum of the distances of the worst users. The reasoning of this assignment is to give more importance to the distances of worst users, since that set is the one that do not fulfill the requirement in the OF case. Similarly, in the UF case, the assignment using the $\max()$ function is for the worst users, which are the users that do not fulfill the requirement in such a case. 

The sub-state $s_u[n]$ is comprised of $\scriptsize s_u[n] = \Big( \hat{\mathbb{E}}\left[\mathbf{\tilde{W}}[n]\right], \hat{\mathbb{S}}\left[\mathbf{\tilde{W}}[n]\right], \hat{\mathbb{E}}\Big[\mathbf{I}_\mathrm{MCS}[n]\Big], \hat{\mathbb{S}}\Big[\mathbf{I}_\mathrm{MCS}[n]\Big] \Big) $, where $\scriptsize \hat{\mathbb{E}}\left[\bullet \right]$ and $\scriptsize \hat{\mathbb{S}}\left[\bullet \right]$ stands for the empirical mean and standard deviation respectively. $\mathbf{\tilde{W}}[n]$ represents the vector of normalized delays of all users and 
$\scriptsize \mathbf{I}_\mathrm{MCS}[n]= \Big({I}_{\mathrm{MCS},1}[n],..,{I}_{\mathrm{MCS},M}[n] \Big)$ is the vector of the MCS indexes reported by all users.


We have considered a discrete action space $\mathcal{A} = \Big\{0,\pm 10^{-4}, \pm 10^{-3}, \pm 10^{-2}, \pm 5\cdot10 ^ {- 2}, \pm 10 ^ {-1}\Big\}$. At each time step $n$, the action taken by the agent will select the step size $\Delta \beta [n] \in \mathcal{A}$ that maximizes the expected cumulative reward. 

\subsection{Reward Function}
%
The proposed reward function encourages the agent to stay in the FF case by taking into account two aspects: (a) the fairness state; and (b) the action taken by the agent. The value of the reward function is defined in (\ref{eq:reward_general}), (\ref{eq:reward_uf}) and (\ref{eq:reward_of}):
\vspace{-2mm}
\begin{equation}
\small
\label{eq:reward_general}
    {r}[n+1] = 
    \begin{cases}
      {r}_{\mathrm{UF}}[n+1] & \mathrm{if} \;  \mathcal{C}[n] = \mathrm{UF}\\
      1 & \mathrm{if} \;  \mathcal{C}[n] = \mathrm{FF}\\
      {r}_{\mathrm{OF}}[n+1] & \mathrm{if} \;  \mathcal{C}[n] = \mathrm{OF}
    \end{cases}
\end{equation}
\vspace{-2mm}
\begin{equation}
\small
\label{eq:reward_uf}
  {r}_{\mathrm{UF}}[n+1] = 
    \begin{cases}
      \Delta\beta[n] & \mathrm{if} \;  \Delta \beta[n] > 0\\
      -1 & \mathrm{if} \; \Delta\beta[n] \le 0
    \end{cases}
\end{equation}
\vspace{-2mm}
\begin{equation}
\small
\label{eq:reward_of}
  {r}_{\mathrm{OF}}[n+1] = 
    \begin{cases}
      -\Delta\beta[n] & \mathrm{if} \; \Delta \beta[n] < 0\\
      -1 & \mathrm{if} \; \Delta\beta[n] \ge 0
    \end{cases}
\end{equation}
The reasoning of \eqref{eq:reward_general} is to encourage the agent to stay in FF state with a maximal positive reward (i.e., $1$). If the fairness case is UF, the agent should increase the $\beta$ parameter to augment the fairness as per remark \ref{rem:beta behaviour}. Thus, according to \eqref{eq:reward_uf}, the reward is positive if $\Delta \beta[n]$ is positive, which means that the action made in previous instant was adequate; otherwise the rewards in negative to penalize the action of decreasing $\beta$ that was made through a negative $\Delta \beta[n]$. Analogously, for OF case the agent is encouraged to decrease the fairness through the reward function as given with \eqref{eq:reward_of}. 

\section{Numerical Results and Discussions}
 \label{sec:results}
 

To simulate the environment we have assumed a single cell scenario with $M =$ 60 active users. 
A detailed list of the network parameters is shown in Table \ref{tab:net_param}.


\begin{table}[t]
\scriptsize
\caption{Network parameters setting}
\label{tab:net_param}
\begin{tabular}{|l|l|l|l|}
\noalign{\hrule height 1pt}
\textbf{Parameter} & \textbf{Value} & \textbf{Parameter} & \textbf{Value} \\ 
\noalign{\hrule height 1pt}
 $N$                   & $100$ RBs     &      Carrier frequency (GHz)                & $5$                       \\
User speed  (Km/h)                     & $5$         & Delay Spread ($\mu s$)                      & $100$                      \\
TTI (ms) & $1$                   & Number of users                  & $60$                          \\
$S_\mathrm{CBR}$ (bytes)                    &  $850$           & $T_\mathrm{CBR}$ (ms)                    &  $6$                             \\
$\mu_\gamma$ (dB)                & $15$  & $\sigma_\gamma$ (dB) & $3$  \\
$\mathrm{BLER}_T$           & $0.1$                   & PF window size ($T_{PF}$)                & $100$ \\ 
$\delta_u$ & 0.05& $T_u$ (ms) & $100$\\
\noalign{\hrule height 0.5pt}
\end{tabular}
\end{table}

The agent implements a DQL with a  neural network comprised of $L$ layers, being $N_\ell$ the number of nodes configured for each layer where  $\ell \in [1,L]$. It should be noted  that the number of nodes in both the input and output layers is fixed, corresponding to the dimensions of the state and action spaces respectively. Based on \cite{Comsa19}, the configuration $\{L = 3, N_2 = 60\}$ is selected to find a balance between flexibility and complexity of our learning system. 
%
%
The agent is trained with a Decayed $\epsilon$-greedy policy. During the training stage the agent was trained during $2\cdot 10^5$ steps (i.e. TTIs). 

\begin{figure}[t]
\centering
\includegraphics[width=0.71\columnwidth]{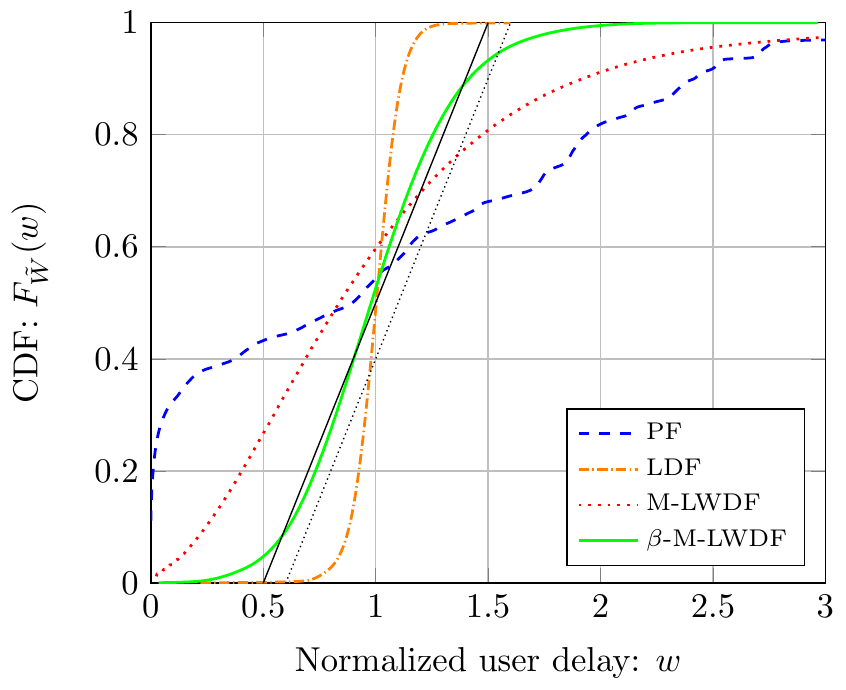}
\caption{Comparison between the CDF of the normalized user delay of different resource allocation algorithms. The $\beta$-M-LWDF considers $\lambda=20\%$ of the users as best users and $\psi=10\%$ of outliers.}
\label{fig:cdf}
\end{figure}

Fig. \ref{fig:cdf} shows the CDF of the normalized user delay for different scheduling policies. 
It can be observed that PF provides the most unfair results as its utility function does not consider the delay. On the contrary, LDF gives strict priority to the delay, thus showing a clear over-fair behaviour, since the best users tend to have delays close to the average delay (i.e., unit normalized delay). M-LWDF provides intermediate results although it still presents an unfair behaviour, since the worst users tend to have high normalized delays in statistical terms. Finally, our proposed algorithm, $\beta$-M-LWDF, is able to fulfill the delay requirement for most of the users. 

\begin{table}[t]
\scriptsize
\caption{Average Delay results of different resource allocation algorithms}
\label{tab:av_delay}
\begin{tabular}{|l|l|l|}
\noalign{\hrule height 1pt}
\textbf{Algorithm} & \textbf{Average delay (ms)} & \textbf{Max. average delay (ms)}  \\ 
\noalign{\hrule height 1pt}
    LDF & $163.2$ & $4192.7$ \\
    M-LWDF & $37.3$ & $134.0$ \\
    $\beta$-M-LWDF & $53.6$ & $95.0$ \\
    PF & $1228.3$ & $3964.5$ \\
\noalign{\hrule height 0.5pt}
\end{tabular}
\end{table}

%
In table \ref{tab:av_delay} it is shown the average delay, which is averaged in time and user domains, and the maximum average delay on time domain. 
%
%
It is observed that LDF leads to the highest delay fairness, but also to the highest average delay.  This is due to the over-fair behaviour of such algorithm. 
It is observed that M-LWDF achieves the smallest average delay. Nevertheless, the average delay of the worst user is clearly greater than with our $\beta$-M-LWDF algorithm. 
%

\begin{figure}[t]
\centering
\includegraphics[width=0.71\columnwidth]{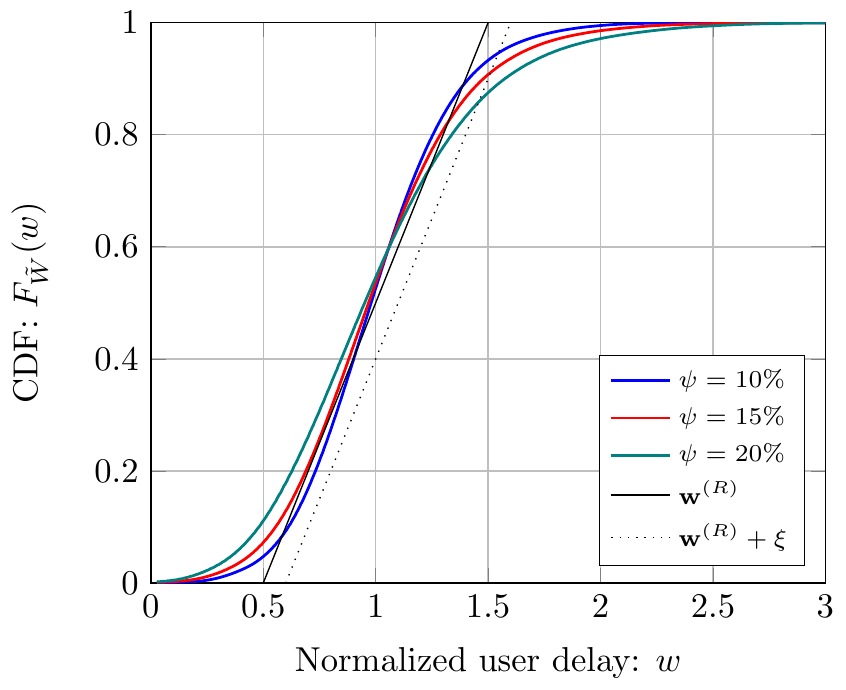}
\caption{CDF of the normalized delay with $\beta$-M-LWDF for different outliers percentages, $\psi=\{10 \%, 15 \%, 20 \%\}$.}
\label{fig:cdf_xi}
\end{figure}

\begin{table}[t]
\scriptsize
\caption{Percentage of time in different fairness cases}
\label{tab:percentage}
\begin{tabular}{|l|l|l|l|l|}
\noalign{\hrule height 1pt}
\textbf{$\psi$} & \textbf{Av. delay (ms)} & FF time ($\%$) & UF time ($\%$) & OF time ($\%$) \\ 
\noalign{\hrule height 1pt}
    $10\%$ & $53.60$ & $90.47$ & $6.60$ & $2.93$ \\
    $15\%$ & $47.03$ & $86.68$ & $12.91$ & $0.41$ \\
    $20\%$ & $39.76$ & $87,17$ & $12.83$ & $0.0$ \\
\noalign{\hrule height 0.5pt}
\end{tabular}
\end{table}

Fig. \ref{fig:cdf_xi} shows the CDF of normalized delay with $\beta$-M-LWDF algorithm for different outliers percentages, $\psi$. It can be observed that lower $\psi$ values increase the delay fairness since the CDF tend to be more centered at the unit normalized delay ($w=1$). Nevertheless, reducing $\psi$ also tends to increase the average delay as shown with table \ref{tab:percentage}. Such a table represents the percentage of time that the system is on each of the fairness cases. It can be observed that the proposed algorithm achieves a high percentage of time on the desired FF case. 
\vspace{-1mm}
\section{Conclusions}
 \label{sec:conclusions}
We have proposed a novel framework based on deep RL to provide an adequate latency fairness. Our proposal includes a new scheduling policy, named as $\beta$-M-LWDF, which is able to adjust instantaneously the allocation criteria based on the experienced delay of the users at each TTI. Simulation results show that our proposal ourperforms other well known scheduling solutions like PF, LDF or M-LWDF in terms of latency fairness and average delay. 

\appendices
\section{}\label{proof:remark} 
When $\beta \to \infty$, \eqref{eq:bmlwdf} can be expressed as follows
\begin{equation}
\small
\arg \mathop {\max}\limits_{u,k} \left\{ \lim\limits_{\beta \to \infty} {\log(g_u)  + \beta \log \left(W_u [n]\right) + \log \left(r_{u,k} [n]\right)} \right\}, 
\end{equation}
\noindent where it has be used the fact that any strictly monotonic function does not change the result of the $\arg \max$ operator. Finally, the proof is completed after applying the following two facts: (i)  the limit when $\beta \to \infty$ only depends on the term multiplied by $\beta$, and (ii) any positive scalar that multiplies a function does not change the result of the $\arg \max$ operator.
%

\end{document}